\documentclass{article}
\usepackage{amsmath, amssymb, amsfonts}
\title{A regular version of the extremal RN spacetime}  
\author{Hristu Culetu, \\Ovidius University, Dept. of Physics and Electronics, \\ Mamaia Avenue 124, 900527 Constanta, Romania, \footnote{e-mail : hculetu@yahoo.com}}

\begin{document}
\numberwithin{equation}{section}
\pagenumbering{arabic}
\maketitle
\newcommand{\fv}{\boldsymbol{f}}
\newcommand{\tv}{\boldsymbol{t}}
\newcommand{\gv}{\boldsymbol{g}}
\newcommand{\OV}{\boldsymbol{O}}
\newcommand{\wv}{\boldsymbol{w}}
\newcommand{\WV}{\boldsymbol{W}}
\newcommand{\NV}{\boldsymbol{N}}
\newcommand{\hv}{\boldsymbol{h}}
\newcommand{\yv}{\boldsymbol{y}}
\newcommand{\RE}{\textrm{Re}}
\newcommand{\IM}{\textrm{Im}}
\newcommand{\rot}{\textrm{rot}}
\newcommand{\dv}{\boldsymbol{d}}
\newcommand{\grad}{\textrm{grad}}
\newcommand{\Tr}{\textrm{Tr}}
\newcommand{\ua}{\uparrow}
\newcommand{\da}{\downarrow}
\newcommand{\ct}{\textrm{const}}
\newcommand{\xv}{\boldsymbol{x}}
\newcommand{\mv}{\boldsymbol{m}}
\newcommand{\rv}{\boldsymbol{r}}
\newcommand{\kv}{\boldsymbol{k}}
\newcommand{\VE}{\boldsymbol{V}}
\newcommand{\sv}{\boldsymbol{s}}
\newcommand{\RV}{\boldsymbol{R}}
\newcommand{\pv}{\boldsymbol{p}}
\newcommand{\PV}{\boldsymbol{P}}
\newcommand{\EV}{\boldsymbol{E}}
\newcommand{\DV}{\boldsymbol{D}}
\newcommand{\BV}{\boldsymbol{B}}
\newcommand{\HV}{\boldsymbol{H}}
\newcommand{\MV}{\boldsymbol{M}}
\newcommand{\be}{\begin{equation}}
\newcommand{\ee}{\end{equation}}
\newcommand{\ba}{\begin{eqnarray}}
\newcommand{\ea}{\end{eqnarray}}
\newcommand{\bq}{\begin{eqnarray*}}
\newcommand{\eq}{\end{eqnarray*}}
\newcommand{\pa}{\partial}
\newcommand{\f}{\frac}
\newcommand{\FV}{\boldsymbol{F}}
\newcommand{\ve}{\boldsymbol{v}}
\newcommand{\AV}{\boldsymbol{A}}
\newcommand{\jv}{\boldsymbol{j}}
\newcommand{\LV}{\boldsymbol{L}}
\newcommand{\SV}{\boldsymbol{S}}
\newcommand{\av}{\boldsymbol{a}}
\newcommand{\qv}{\boldsymbol{q}}
\newcommand{\QV}{\boldsymbol{Q}}
\newcommand{\ev}{\boldsymbol{e}}
\newcommand{\uv}{\boldsymbol{u}}
\newcommand{\KV}{\boldsymbol{K}}
\newcommand{\ro}{\boldsymbol{\rho}}
\newcommand{\si}{\boldsymbol{\sigma}}
\newcommand{\thv}{\boldsymbol{\theta}}
\newcommand{\bv}{\boldsymbol{b}}
\newcommand{\JV}{\boldsymbol{J}}
\newcommand{\nv}{\boldsymbol{n}}
\newcommand{\lv}{\boldsymbol{l}}
\newcommand{\om}{\boldsymbol{\omega}}
\newcommand{\Om}{\boldsymbol{\Omega}}
\newcommand{\Piv}{\boldsymbol{\Pi}}
\newcommand{\UV}{\boldsymbol{U}}
\newcommand{\iv}{\boldsymbol{i}}
\newcommand{\nuv}{\boldsymbol{\nu}}
\newcommand{\muv}{\boldsymbol{\mu}}
\newcommand{\lm}{\boldsymbol{\lambda}}
\newcommand{\Lm}{\boldsymbol{\Lambda}}
\newcommand{\opsi}{\overline{\psi}}
\renewcommand{\tan}{\textrm{tg}}
\renewcommand{\cot}{\textrm{ctg}}
\renewcommand{\sinh}{\textrm{sh}}
\renewcommand{\cosh}{\textrm{ch}}
\renewcommand{\tanh}{\textrm{th}}
\renewcommand{\coth}{\textrm{cth}}

\begin{abstract}
A modified extremal Reissner-Nordstrom geometry, void of singularities, is proposed in this work, by means of an exponential factor depending on a positive constant $k$. All the metric coefficients are positive and finite and the spacetime has no any horizon. The curvature invariants are regular at the origin of coordinates and at infinity. The energy conditions for the stress tensor associate to the imperfect fluid are investigated. The gravitational field presents repulsive properties near the gravitational radius associated to the mass $m$. With the choice $k = 1/m$, the Komar energy $W_{K}$ of the mass $m$ changes its sign at $r = \lambda$ ($\lambda$ is the Compton wavelength of $m$), when the classical energy $mc^{2}$ equals the energy $\hbar c/r$.
 \end{abstract}

 \section{Introduction}
 Extremal black holes are distinguished from non-extremal ones, even at the classical level. This is in accordance with the view that extremal blach holes (EBHs) may be considered as solitons to be formed quantum-mechanically by pair production (and whose entropy would be thought to vanish) \cite{CJR}. If so, some properties of EBHs might be different from near-extremal ones \cite{GK}. 

 Carroll et al. \cite{CJR} investigated also the entropy of EBHs. Keeping in mind that the surface gravity and the temperature vanish \cite{NC}, the entropy of an EBH also vanishes \cite{EC} (Edery and Constantineau justified that due to the time-independent geometry throughout, which corresponds to a single classical microstate). It is worth noting that, near the horizon, the EBH has an $AdS_{2} \times S^{2}$ structure. The entropy of the $AdS_{2} \times S^{2}$ compactification solution does not vanish \cite{CJR}. However, the entropy calculation concerns the near-horizon region and not the original black hole solution. Therefore, Carroll et al. concluded that the EBHs should have zero entropy.

Bonanno and Reuter \cite{BR} argued that the renormalization group improved Schwarzschild spacetime is similar to a Reissner-Nordstrom (RN) black hole and the critical quantum black hole mass $M_{cr}$ is of the order of the Planck mass. In their view, the Hawking evaporation process is ''switched off'' once the mass approaches  $M_{cr}$. They also noticed that the near-horizon geometry of the critical BH is, to leading order, the Robinson-Bertotti line element for the product of a two-dimensional AdS with a two-sphere, $AdS_{2} \times S^{2}$. 
  Horowitz et al. \cite{HKS} studied the near-horizon geometry of a four-dimensional extremal black hole. When the cosmological constant $\Lambda$ is negative, they showed that the tidal forces on infalling particles diverge when the horizon is crossed, albeit all scalar curvature invariants remain finite. Similar effects appear when $\Lambda$ is positive, but not when it vanishes.
	
  Non-singular BHs in general have been firstly studied by Bardeen \cite{JB}, who proposed a model of charged matter collapse with a charged matter core inside the
black hole, with no central singularity. Hayward \cite{SH} considered a spherically-symmetric regular BH with, apart from $m$, an extra parameter $l$. His metric has a de Sitter (deS) form when $r = 0$ is approached and becomes Schwarzschild for $l = 0$. More recently, Frolov \cite{VF} discussed useful generalizations of the Hayward spacetime (with finite curvature invariants), in four-dimensional and higher dimensional geometries. The additional parameter $l$ determines the scale where modification of the solution of the Einstein equations becomes significant. Simpson and Visser \cite{SV}, using a special exponential function (see also \cite{HC1, HC2}) and the increasingly important Lambert $W$-function, avoid rather messy cubic and quartic polynomial equations used in the previous models.  
	
	We propose in this work a modification of the RN extremal BH for rendering it void of singularities. For that goal one introduces an exponential factor in the metric components. The new geometry is regular throughout and all scalar invariants are finite at the origin of coordinates and at infinity. The timelike Killing vector is vanishing nowhere and the geometry has no any horizon, so the name ''black hole'' is no longer appropriate. The spacetime can be considered in its own right even though we started with a RN black hole.We investigated the properties of the energy-momentum tensor acting as the source of the modified metric. We also found that the gravitational field presents repulsive properties close to the origin of coordinates, where the source is located. Moreover, the tidal forces on free falling particles are not divergent, thanks to the finiteness of all the components of the Riemann tensor.
	
	Throughout the work geometrical units $G = c = \hbar = 1$ are used, unless otherwise specified.
	
  \section{Modified extremal BH}
	Let us consider the Reissner-Nordstrom geometry
  \begin{equation}
   ds^{2} = -\left(1 - \frac{2m}{r} +\frac{q^{2}}{r^{2}}\right) dt^{2} + \frac{dr^{2}}{\left(1 - \frac{2m}{r} + \frac{q^{2}}{r^{2}}\right)} + r^{2} d \Omega^{2},         
 \label{2.1}
 \end{equation}
where $d \Omega^{2}$ stands for the metric on the unit two-sphere and $m$ and $q$ are the mass and, respectively, the charge of the BH. As is well-known, when $m^{2}>q^{2}$, the RN black hole has two horizons: one at $r_{+} = m + \sqrt{m^{2} - q^{2}}$ (event horizon) and $r_{-} = m - \sqrt{m^{2} - q^{2}}$ (Cauchy horizon). If $m^{2}<q^{2}$, there is no any horizon and the surface $r = 0$ becomes a naked singularity. The case $m^{2} = q^{2}$ represents an extremal BH, when the line-element (2.1) acquires the form
  \begin{equation}
   ds^{2} = -\left(1 - \frac{m}{r}\right)^{2} dt^{2} + \frac{dr^{2}}{\left(1 - \frac{m}{r}\right)^{2}} + r^{2} d \Omega^{2},         
 \label{2.2}
 \end{equation}
which is time independent even inside the BH horizon $r = m$. That is a consequence of the fact that the timelike Killing vector remains timelike everywhere (there is no a $r-t$ signature flip when the event horizon $r = m$ is crossed). However, the true singularity at the origin $r = 0$ survives when the BH becomes extremal. 

 To get rid of the singularity, we propose a modified version of the metric (2.1) by means of an exponential factor, by analogy with \cite{HC1}. The modified spacetime is given by
  \begin{equation}
   ds^{2} = -\left[1 - \left(\frac{2m}{r} -\frac{m^{2}}{r^{2}}\right)e^{-\frac{k}{r}}\right] dt^{2} + \frac{dr^{2}}{1 - \left(\frac{2m}{r} - \frac{m^{2}}{r^{2}}\right)e^{-\frac{k}{r}}} + r^{2} d \Omega^{2},         
 \label{2.3}
 \end{equation} 
with $k$ a positive constant and we replaced the charge $q$ with $m$. We take the geometry (2.3) in its own right, not mandatory related to the RN geometry. It is clear that (2.3) becomes the extremal BH (2.2) if we take $k = 0$ (or $r>>k)$, when the exponential factor is approximated to unity). The next step is to search the properties of the metric function 
  \begin{equation}
	f(r) \equiv -g_{tt} = 1 - \left(\frac{2m}{r} -\frac{m^{2}}{r^{2}}\right)e^{-\frac{k}{r}}
 \label{2.4}
 \end{equation} 
It tends to unity when $r \rightarrow 0,~ r \rightarrow \infty$ or $r = m/2$, and has two extrema: a maximum at $r_{1} = (m + k - \sqrt{m^{2} + k^{2}})/2$ and a minimum at $r_{2} = (m + k + \sqrt{m^{2} + k^{2}})/2$. Noting that $f(r)$ is always positive, because $(2m/r - m^{2}/r^{2})$ is smaller than unity for any $r>0$.

As an example, consider a particular situation with $k = m$. In that case, the maximum is at $r_{1} = m(1 - \sqrt{2}/2) \approx 0.3m$, with $f(r_{1}) = 1.16$, and the minimum is at $r_{2} = m(1 + \sqrt{2}/2) \approx 1.7m$, with $f(r_{2}) = 0.412 >0$. Therefore, $f(r)$ is positive for any $r$, as we already noticed. It is worth observing that the extremal RN horizon at $r = m$ is somewhere between  $r_{1}$ and $r_{2}$. Being $f(r)$ always positive, the metric (2.3) has no any horizon and is regular at the origin of coordinates, where it becomes Minkowskian. 

   We now consider a static observer in the geometry (2.3) having a velocity vector field
	  \begin{equation}
  u^{b} = \left(\frac{1}{\sqrt{1 - \left(\frac{2m}{r} - \frac{m^{2}}{r^{2}}\right)e^{-\frac{k}{r}}}}, 0, 0, 0 \right),~~~ u^{b}u_{b} = -1,  
 \label{2.5}
 \end{equation} 
where the Latin indices take the values $(t, r, \theta, \phi)$. The only nonzero component of the covariant acceleration $a^{b}= u^{a}\nabla_{a}u^{b}$ is
	  \begin{equation}
  a^{r} = \frac{m}{r^{2}}\left(1 - \frac{m}{r} - \frac{k}{r} + \frac{km}{r^{2}}\right)e^{-\frac{k}{r}} = \frac{1}{2}f'(r),
 \label{2.6}
 \end{equation} 
where $f'(r) = df(r)/dr$. The radial acceleration is vanishing when $r \rightarrow 0$ and at infinity. If $r>>k$, we get $a^{r} \approx (m/r^{2})(1 - m/r)$, namely the value corresponding to the extremal BH. Moreover, $a^{r}$ vanishes where $f'(r) = 0$, i.e. at the values of $r$ where $f(r)$ has its extrema, namely at the previous $r_{1}$ and $r_{2}$, being negative for $r \in (r_{1},r_{2})$. This means the gravitational field felt by a static observer is repulsive in the aforementioned domain of $r$. 

 As long as the curvature invariants (the scalar curvature and the Kretschmann scalar) for the metric (2.3) are concerned, one obtains
	  \begin{equation}
  R^{a}_{~a} = \frac{2km^{2}}{r^{5}}\left(1 + \frac{k}{m} - \frac{k}{2r}\right)e^{-\frac{k}{r}}
	\label{2.7}
	\end{equation}
	and
	\begin{equation}
	\begin{split}
	 K = \frac{48m^{2}}{r^{6}}(1 - \frac{2(k + m)}{r} 
	 + \frac{12k^{2} + 24km + 7m^{2}}{6r^{2}}  
	 - \frac{k(4k^{2} + 18km + 11m^{2})}{6r^{3}}\\ + \frac{k^{2}(k^{2} + 10km + 13m^{2})}{12r^{4}} - \frac{k^{3}m(k + 3m)}{12r^{5}} + \frac{k^{4}m^{2}}{48r^{6}})e^{-\frac{k}{r}}
 \label{2.8}
\end{split}
 \end{equation} 
It is clear from (2.7) and (2.8) that the two invariants are vanishing if $r \rightarrow 0$ or $r \rightarrow \infty$. Moreover, if $k = 0$, one obtains
	\begin{equation}
	 K = \frac{48m^{2}}{r^{6}}\left(1 - \frac{2m}{r} + \frac{7m^{2}}{6r^{2}}\right),  
 \label{2.9}
 \end{equation} 
a value corresponding to the standard extremal RN black hole.

\section{Stress tensor properties}
We look now for the source of curvature of the geometry (2.3), namely we need the components of the stress tensor to be inserted on the r.h.s. of Einstein's equations $G_{ab} = 8\pi T_{ab}$ for to get (2.3) as an exact solution. The only nonzero components of the Einstein tensor are given by
  \begin{equation}
	\begin{split}
	G^{t}_{~t} = G^{r}_{~r} = -\frac{m^{2}}{r^{4}}\left(1 + \frac{2k}{m} - \frac{k}{r}\right)e^{-\frac{k}{r}},~~~G^{\theta}_{~\theta} = G^{\phi}_{~\phi}\\ = \frac{m(m + 2k)}{r^{4}}\left[1 - \frac{k(k + 2m)}{(m + 2k)r} + \frac{mk^{2}}{2(m + 2k)r^{2}}\right]e^{-\frac{k}{r}}.
 \label{3.1}
\end{split}
 \end{equation}
As the source of the geometry (2.3) we employ an energy-momentum tensor corresponding to an imperfect fluid \cite{GRM, HC3, HC4}
	  \begin{equation}
	 T_{ab} = (p_{t} + \rho) u_{a} u_{b} + p_{t} g_{ab} + (p_{r} - p_{t}) n_{a}n_{b},
 \label{3.2}
 \end{equation}
where $\rho$ is the energy density of the fluid, $p_{r}$ is the radial pressure, $p_{t}$ are the transversal pressures and $n^{a} = (0, \sqrt{f(r)}, 0, 0)$ is a vector ortogonal to $u^{a}$, with $u^{a}n_{a} = 0$ and $n^{a}n_{a} = 1$. Using the ansatz (2.5) for $u^{a}$, the Einstein equations and the form (3.2) of the stress tensor, one finds that
 	  \begin{equation}
		\begin{split}
		8\pi \rho = 8\pi T^{a}_{~b}u^{b}u_{a} = \frac{m^{2}}{r^{4}}\left(1 + \frac{2k}{m} - \frac{k}{r}\right)e^{-\frac{k}{r}} = -8\pi p_{r},\\ 8\pi p_{t} = \frac{m^{2}}{r^{4}}\left(1 + \frac{2k}{m} - \frac{k^{2}}{mr} - \frac{2k}{r} + \frac{k^{2}}{2r^{2}} \right)e^{-\frac{k}{r}}.
 \label{3.3}
\end{split}
 \end{equation}
One observes that $\rho$ is negative for $r<km/(m + 2k)$, positive for $r> km/(m + 2k)$ and is vanishing when $r \rightarrow 0$, $r \rightarrow \infty$ and at $r = km/(m + 2k)$. In addition, the choice $k = 0$ leads to the well-known expressions $\rho = -p_{r} = p_{t} = m^{2}/8\pi r^{4}$. For the particular case $k = m$, $\rho$ has a minimum $8\pi \rho_{min} = -3888/e^{6}m^{2}$ at $\bar{r}_{1} = m/6$, a maximum $8\pi \rho_{max} = 16/e^{2}m^{2}$ at $\bar{r}_{2} = m/2$ and is vanishing at $\bar{r_{0}} = m/3$. Note that the two extrema of $\rho$ are located below the event horizon $r = m$ of the RN extremal BH.

 As Simpson and Visser \cite{SV} have noticed, to examine where the energy density is maximised is of interest, due to the exponential supression of the mass. They found that their $\rho$ is maximised at $r = a/4$, where $a$ plays the same role as $k (= m)$ in our situation. Our $\rho$ is maximised at $\bar{r_{2}} = m/2$, a double value compared to theirs (anyway, of the same order of magnitude). In addition, when $a>2m/e$, their geometry has no horizons, which means no BH, a situation similar to ours. Another interesting analogy of our metric (2.3) with that of Simpson and Visser is the non-standard asymptotically Minkowskian core, when $r \rightarrow 0$. As we shall see, what is new (to the best of our knowledge) is the regular geometry (2.3), the fact that it has no horizons and the application of our model in microphysics (see Sec.5). In addition, as a new task compared to the previous studies on the non-singular BHs, we computed the Komar energy, both for $k = m$ and $k = 1/m$ (in Sec. 4 and 5), and studied its interesting properties, especially the competition between the classical and quantum contribution for the application in microphysics. 

 As far as the transversal pressures are concerned, they are negative for $r\in (r_{1}^{*}, r_{2}^{*})$, with $r_{1,2}^{*} = k(k + 2m \mp \sqrt{k^{2} + 2m^{2}})/2(m + 2k)$ and positive outside this interval. 

  Let us check now the energy conditions corresponding to the stress tensor (3.2).
	
	- WEC (weak energy condition, $\rho\geq 0,~~\rho + p_{r}\geq 0,~~ \rho + p_{t}\geq 0$)
	
	  We already found that $\rho\geq 0$ when $r\geq km/(m + 2k)$. The condition $\rho + p_{t}\geq 0$ gives us $r\in (0, r'_{1}]\cup [r'_{2}, \infty)$, where $r'_{1,2} = k(k + 3m \mp \sqrt{(k - m)^{2} + 4m^{2}})/4(m + 2k)$. The WEC is obeyed when $r$ is within the above domains.
		
	- NEC (null energy condition, $\rho + p_{r}\geq 0,~~ \rho + p_{t}\geq 0$)
	
	   NEC is included in WEC, hence we have $r\in (0, r'_{1}]\cup [r'_{2}, \infty)$.
		
	- SEC (strong energy condition, NEC and $\rho + p_{r} + 2p_{t}\geq 0$)
	
	   The new condition leads to $p_{t}\geq 0$. That means $r\in (0, r_{1}^{*}]\cup [r_{2}^{*}, \infty)$.
		
	- DEC (dominant energy condition, $\rho \geq |p_{r}|, \rho \geq |p_{t}|$)
	
	  The first condition is satisfied if $\rho\geq 0$, namely $r\geq km/(2k + m)$. For the second we get two other conditions: $r\geq km/2(k + m)$ and $-\rho \leq p_{t}\leq \rho$, which leads to  $r\in (0, r'_{1}]\cup [r'_{2}, \infty)$. Noting that the inequality $r\geq km/2(k + m)$ is not necessary, being included in $r\geq km/(2k + m)$.
		
		Let us make now a first choice of the length $k$. We have only one constant in the model, with units of distance: the source mass $m$, or more precisely, the half-gravitational radius of it. This is clearly appropriate for macroscopic masses, that is for masses $m>m_{P} = 10^{-5}$gr, where $m_{P}$ is the Planck mass. We consider Planck's mass to be the boundary between the macroscopic and microscopic masses.
		
		The energy density, the radial pressure and the transversal pressures acquire the form
 \begin{equation}
\rho = -p_{r} = \frac{3m^{2}}{r^{4}}\left(1 - \frac{m}{3r}\right)e^{-\frac{m}{r}},~~~p_{t} = \frac{3m^{2}}{r^{4}}\left(1 - \frac{m}{r} + \frac{m^{2}}{6r^{2}}\right)e^{-\frac{m}{r}}.
 \label{3.4}
 \end{equation} 
	We see that the $\rho$ and $p_{t}$ expressions given above contain few correcting terms compared to those from \cite{HC1} (Eqs.3.1). Therefore, we do not intend to show more details about the properties of the above quantities. Concerning the energy conditions, it can be shown that, with $k = m$, the WEC, NEC, SEC and DEC are observed provided $r\geq r_{2}^{*} = (m/2)(1 + \sqrt{3}/3) \approx 0.78m$, which is smaller than the standard gravitational radius of the mass $m$.	
	
	\section{Energetic considerations}
	It is instructive to compute the Komar mass-energy $W_{K}$ \cite{TP1}
 	\begin{equation}
	W_{K} = 2 \int_{V}(T_{ab} - \frac{1}{2} g_{ab}T^{c}_{~c})u^{a} u^{b} N\sqrt{h} d^{3}x ,
 \label{4.1}
 \end{equation} 
where $V$ is a 3-volume in our static metric, $N^{2} = -g_{tt}$ is the lapse function, $h = r^{4}sin^{2}\theta /f(r)$ is the determinant of the spatial 3 - metric, $h_{ab} = g_{ab} + u_{a} u_{b}$, and $d^{3}x = dr~d\theta~d\phi$. With $u^{a}$ from (2.5) and $T_{ab}$ from (3.2), we get
 	\begin{equation}
	(T_{ab} - \frac{1}{2} g_{ab}T^{c}_{~c})u^{a} u^{b} = p_{t},~~~ N\sqrt{h} = r^{2}sin\theta .
 \label{4.2}
 \end{equation} 
 Therefore, Eq.(4.1) yields
 	\begin{equation}
	W_{K} =  \int_{0}^{r}\left[\frac{m(m + 2k)}{r'^{2}} - \frac{mk(k + 2m)}{r'^{3}} + \frac{m^{2}k^{2}}{2r'^{4}}\right]e^{-\frac{k}{r'}}dr' ,
 \label{4.3}
 \end{equation} 
 Keeping in mind that
  \begin{equation}
		\int{\frac{1}{r^{3}}e^{-\frac{k}{r}}}dr = \frac{1}{k}\left(\frac{1}{k} + \frac{1}{r}\right)e^{-\frac{k}{r}},~~~ \int{\frac{1}{r^{4}}e^{-\frac{k}{r}}}dr = \frac{2}{k^{3}}\left(1 + \frac{k}{r} + \frac{k^{2}}{2r^{2}}\right)e^{-\frac{k}{r}}.
 \label{4.4}
 \end{equation} 
One obtains
 	\begin{equation}
	W_{K} = \left[\frac{m(m + 2k)}{k} - m(k + 2m)(\frac{1}{k} + \frac{1}{r}) + km^{2}(\frac{1}{k^{2}} + \frac{1}{kr} + \frac{1}{2r^{2}})\right] e^{-\frac{k}{r}} .
 \label{4.5}
 \end{equation} 
After a rearangement of terms, we get
 	\begin{equation}
	W_{K} = m\left(1 - \frac{m}{r} - \frac{k}{r} + \frac{km}{2r^{2}}\right) e^{-\frac{k}{r}} ,
 \label{4.6}
 \end{equation} 
It is worth observing that, when $k = 0$, one obtains $W_{K} = m(1 - m/r) = m - m^{2}/r$, a sum between the rest energy and the proper gravitational energy. In addition, if $r \rightarrow \infty$, (4.6) gives us $W_{K,\infty} = m$, as expected; in contrast, $W_{K}$ is vanishing when $r\rightarrow 0$. In terms of the radial acceleration (2.6), we have $W_{K}(r) = r^{2} a^{r}$ or, written differently, with fundamental constants included, , $a^{r} = GM(r)/r^{2} = G W_{K}(r)/c^{2}r^{2}$. In other words, the radial acceleration at the distance $r$ from the origin depends directly on the Komar mass. The sign of $W_{K}$ is given, of course, by the sign of $a^{r}$ or $f'(r)$ (see below (2.4)). Hence, $W_{K}$ is negative when $r\in (r_{1}, r_{2})$ (where the gravitational field is repulsive) and positive otherwise.

\section{The choice $k = 1/m$}
 This second choice of $k$ leads to a semiclassical model of our system described by the spacetime (2.3): the usage of the Planck constant $\hbar$, such that $k$ equals the reduced Compton wavelength associated to the mass $m$, namely $\lambda = \hbar/mc$. That means our paradigm has to be applied in microphysics, with $m$ signifying the mass of an elementary particle, with, of course, $m<<m_{P}$.

 Our next task is to investigate the properties of some of the quantities analysed by now.

- The radial acceleration acquires now the form
	  \begin{equation}
  a^{r} = \frac{m}{r^{2}}\left(1 - \frac{m}{r} - \frac{\lambda}{r} + \frac{l_{P}^{2}}{r^{2}}\right)e^{-\frac{\lambda}{r}} ,
 \label{5.1}
 \end{equation} 
where $l_{P} = \sqrt{G\hbar /c^{3}}$ is the Planck length. For an elementary particle (say, a neutron), its gravitational radius is $\approx 10^{-52}cm$ and $\lambda \approx 10^{-13}cm$, so that the second and the fourth terms in the parantheses may be neglected w.r.t. the third. Consequently, we can write 
	\begin{equation}
  a^{r} \approx \frac{m}{r^{2}}\left(1 - \frac{\lambda}{r}\right)e^{-\frac{\lambda}{r}},
 \label{5.2}
 \end{equation} 
and the radial acceleration changes its sign at $r = \lambda$. If we calculate it exactly from (5.1) at that value of $r$, again for a neutron, one obtains $a^{r} = -(m/m_{P})^{4}(1/2\lambda) = -5.16\times 10^{-42}cm/s^{2}$, which is indeed nearly zero ($m_{P}$ is the Planck mass). That is a consequence of the fact that, with $k = \lambda$, we have $r_{1} \approx m/2$ and $r_{2} \approx \lambda$, with $a^{r}<0$ for $r\in (r_{1}, r_{2})$ (see below (2.6)). But $ r_{1}\approx 0$, so that, with a good approximation, $a^{r}<0$ for any $r<\lambda$. We have seen that $a^{r}$ is practically zero at $r = \lambda$ but, however, it is not negligible at $r = \lambda /2$, where we have $a^{r}\approx -1.3 \cdot 10^{-4}cm/s^{2}$. 

 Let us check the value of $a^{r}$ for $r>\lambda$, say $r = 3\lambda /2$, using the approximate expression (5.2), by a comparison with the previous $a^{r}$. We have $a^{r}(3\lambda /2) = - a^{r}(\lambda /2) e^{4/3}/27 = 0.18 \cdot 10^{-4} cm/s^{2}$. One observes that the radial acceleration already changed its sign, and the gravitational field becomes attractive. As an extra example, one could find that $a^{r} \approx 1.4\cdot 10^{-8} cm/s^{2}$ at $r = 2\lambda$, an acceptable value.
We see here something similar with the interaction between two molecules, due to the van der Waals forces. When they are approaching, the force between them is attractive but when the distance is decreased, repulsive forces come into play (in our situation, the second particle is a test particle found close to the neutron).

 - The energy density when $k = \lambda$ is given by
\begin{equation}
8\pi \rho = \frac{m^{2}}{r^{4}}\left(1 + \frac{2m_{P}^{2}}{m^{2}} - \frac{\lambda}{r}\right)e^{-\frac{\lambda}{r}}
 \label{5.3}
 \end{equation} 
For instance, at $r = \lambda$, we get $\rho = 2m_{P}^{2}/\pi e\lambda^{4} = \hbar c/4\pi e\lambda^{4}$, which no longer depends on the Newton constant $G$, but only on $\hbar$ and $c$. We already found that $\rho>0$ for any $r\geq km/(m + 2k)$, which becomes now $r\geq l_{P}^{2}/(m + 2\lambda)\approx l_{P}^{2}/ 2\lambda < l_{P}$. In other words, $\rho$ is practically always positive with the above value of $k$. 

By analogy with the macroscopic case, we are interested in the expression of the maximum value of the energy density. Working in the approximation $m<<m_{P}$, we find that $\rho$ has two extrema: one minimum at $r = R_{1} \approx m^{2}\lambda /4m_{P}^{2}$, with negative $\rho_{min}$, but very close to zero due to the exponential factor, and a maximum at $r = R_{2} \approx \lambda /4$, with $\rho_{max} = \rho ( R_{2}) = 32 m_{P}^{2}/\pi e^{4}\lambda^{4}$. We have $\lambda >>l_{P}$, such that the Markov \cite{MM} criterion is satisfied. We notice that Frolov \cite{VF} used Markov's recipe in his paper, as the limiting curvature conjecture (namely, the Plank value of curvature). 

 We proved that all the energy conditions are obeyed if $r\geq r_{2}^{*}$. With the above value of $k$ and keeping in mind that $\lambda >>m$ we get  $r\geq \lambda/2$. That means for $r
\leq\lambda/2$ some energy conditions are not satisfied.

- The Komar energy acquires the form, with the new value of $k$
    	\begin{equation}
	W_{K}(r) = m\left(1 - \frac{m}{r} - \frac{\lambda}{r} + \frac{l_{P}^{2}}{2r^{2}}\right) e^{-\frac{\lambda}{r}}.
 \label{5.4}
 \end{equation} 
But the second and the fourth terms inside the parantheses may be neglected w.r.t. the third. Therefore
    	\begin{equation}
	W_{K}(r) \approx m\left(1 - \frac{\lambda}{r}\right) e^{-\frac{\lambda}{r}},
 \label{5.5}
 \end{equation} 
which changes sign at $r = \lambda$, as the radial acceleration. When the fundamental constants are inserted, (5.5) yields
    	\begin{equation}
	W_{K}(r) = \left(mc^{2} - \frac{\hbar c}{r}\right) e^{-\frac{\hbar}{mcr}}.
 \label{5.6}
 \end{equation} 
Firstly one observes that, with the above approximation, $W_{K}$ does not depend on $G$. Classically, when $\hbar \rightarrow 0$ or if $r>>\lambda$, one obtains $W_{K} = mc^{2}$, as expected. We notice that the two contributions, classical and quantum, are opposed, being equal at $r = \lambda$, when $W_{K} = 0$.

\section{Conclusions}
As standard BHs, the extremal black holes are not regular at the origin of coordinates. Making them regular throughout might be an important task. It is exactly what we tried to do in the present paper. For that purpose we made use of an exponential factor rendering the spacetime void of singularities. The metric coefficients being positive, our spacetime has no horizons and the label ''black hole'' is no longer appropriate. 

The properties of the stress tensor of the imperfect fluid are studied, especially the energy conditions. An important property of our physical system is the Komar energy, which has been calculated and discussed. We chose for the constant $k$ two values: $k = m$ for the macroscopic case and $k = 1/m$ for the microscopic case, which is the semiclassical one, due to the Planck constant from the expression of the Compton wavelength.

\end{document}